\documentclass[11pt]{revtex4}

\usepackage{amsmath}
\usepackage[dvips]{graphicx}
\usepackage{dcolumn}
\usepackage{bm}
\usepackage{epsfig}

\begin{document}

\title{Diffusion of Macromolecules across the Nuclear Pore Complex }
\author{Rajarshi Chakrabarti, Ananya Debnath and K. L. Sebastian}

\begin{abstract}
Nuclear pore complexes (NPCs) are very selective filters that monitor the
transport between the cytoplasm and the nucleoplasm. Two models have been
suggested for the plug of the NPC. They are (i) it is a reversible hydrogel
or (ii) it is a polymer brush. We propose a mesoscopic model for the
transport of a protein through the plug, that is general enough to cover
both. The protein stretches the plug and creates a local deformation. The
bubble so created (prtoein+deformation) executes random walk in the plug. We
find that for faster relaxation of the gel, the diffusion of the bubble is
greater. Further, on using parameters appropriate for the brush, we find
that the diffusion coefficient is much lower. Hence the gel model seems to
be more likely explanation for the workings of the plug.
\end{abstract}

\affiliation{Department of Inorganic and Physical Chemistry, Indian Institute of Science,
Bangalore 560012, India}
\maketitle

The nuclear envelope in all eucaryotes is perforated with nuclear pores \cite%
{gorlich, wente2, aebi, reiner, wente, lim2}. Each pore has a selective
filter, referred to as the nuclear pore complex (NPC). The NPC is a
self-assembled, eightfold symmetric ringlike structure consisting of eight
copies each of 30-50 different proteins, connecting the inner and outer
nuclear membranes. It regulates the import and export traffic of proteins
and has two distinct modes of transport: passive and facilitated. Passive
transport is non-specific and takes place by ordinary diffusion. Colloidal
gold particles with radii up to 4 nm, and generic proteins up to 50 kDa in
mass, pass efficiently through the NPC in this way \cite{paine}. In
contrast, facilitated translocation allows the passage of objects as large
as several megadaltons. Proteins having a short amino acid sequence known as
nuclear localization signal (NLS) form a complex with transportin \cite%
{wente} (a transporter protein rich in hydrophobic units) and are
transported in this mode. The transport requires specific interactions
between the translocating species and constituents of the NPC and
consequently is highly selective. Gold particles of up to 32-36 nm in
diameter are able to pass through some NPC if they are coated with
nucleoplasmin-importin complexes \cite{aebi}. This suggests that the
interaction of the protein-transportin complex with NPC is essential for
transport. \ Passage of proteins through the NPC has attracted considerable
experimental and theoretical attention \cite{ribbeck,yang,
bickel,kustanovich}. \ According to Ribbeck and G\"{o}rlich (RG) \cite%
{ribbeck} the central plug of the nuclear pore is made of long diblock
copolymers rich in hydrophobic phenylalanine-glycine (FG) units forming a
meshwork. \ Only the macromolecules which can form hydrophobic contacts with
the FG units are incorporated into this network and get transported. \ In an
interesting paper, Bickel and Bruinsma (BB) \cite{bickel} point out that
such a model would lead to a lower rate of diffusion. \ BB suggest that the
central plug is a reversible polymer gel in a poor solvent and a protein in
it experiences an extra noise (they call it \textquotedblleft chemical
noise") arising from the fluctuations of the FG contacts and this extra
noise enhances the diffusion of the protein within the NPC. \ Single
molecule fluorescence microscopy by Yang, Gelles and Musser \cite{yang}
shows that the protein executes random walk inside the central core of the
NPC. \ An alternate model for the plug suggests that it is not a gel but a
polymer brush \cite{Macara2001}. \ \ Surprisingly, there have been
experimental support for both gel and brush models. \ In interesting
experiments, Frey et al. \cite{frey, burke, elbaum} have shown that the
nucleoporins form a hydrogel in vitro, offering support to the model of RG
and BB. \ On the other hand a beautiful study by Lim et al. \cite{LIM} has
found that the proetins, when grafted to a surface behave like an
un-cross-linked brush. \ Also, it is known that the interaction between the
transportin and the FG residues is not just hydrophobic, but involves
hydrogen bonding, electrostatic and van der Waals interactions \cite%
{Isgro2005,Liu2005} and that there are extremely hydrophilic portions in
between the FG units. \

In the following we study a minimalistic model for the transport in the NPC.
Our model is quite general, and would be applicable whether the plug is a
gel or a brush. The actual values of the parameters in the model would
depend on whether it is a gel or a brush. We find that the it is possible
for the protein to diffuse rapidly within a reversible gel, while the
diffusion would be much slower within the brush.

We take the pore complex to be infinitely long (end-effects neglected) and
shall adopt a continuum description for the plug. We use $x$ to denote
position along the direction of the axis of the NPC. \ To make the problem
one dimensional, we imagine the cross section of the pore to be a square,
with width $L_{Y}$ in the $Y$ and height $L_{Z}(=L_{Y})$ in the $Z$
directions. \ We shall assume that the particle has a length $2R_{0}$ in the
$X$ direction, causes a distortion of height $\sim \alpha $ and fills the
pore fully in the $Y$ direction. \ Further, to simplify the analysis, we
assume that periodic boundary conditions are imposed in this direction, with
a period $L_{Y}$. With these, the problem is reduced to two dimensions ($X$
and $Z$). \ The size of the distortion needed to create a cavity to
accommodate the particle will be our important variable. \ Let $\phi (x)$
denote the height of the cavity in the $Z$-direction at the position $x$. \
\ The simplest possible expression for the energy of distortion $E_{dis}$
would have terms quadratic in $\phi (x)$. Thus $E_{dis}=\left( 1/2\right)
\int_{-\infty }^{\infty }dx\left\{ \sigma \left( \partial \phi (x)/\partial
x\right) ^{2}+k\phi (x)^{2}\right\} $. The energy of interaction $%
E_{int}=-E_{b}+\int_{-\infty }^{\infty }dxQ(x-R)\left\{ (k+\Delta
k)/2\left\{ \phi (x)-\alpha \right\} ^{2}-(k/2)\phi (x)^{2}\right\} $ of the
particle at $R$ with the plug can compensate for the distortion energy. \ \ $%
Q(y)$ with $y=x-R$ is a function that determines the interaction between the
particle and the plug. \ \ We assume $Q(y)$ to be a symmetric function of $y$%
, having maximum value $Q(0)=1$. \ Further, we assume $Q(\pm \infty )=0.$\
When the particle enters the plug it would have to break hydrophobic
contacts that may be there, and the energy expenditure for that may be met
by formation of new contacts of the nucleoporins with the particle. All
these together is represented by a single constant $E_{b}$. Thus the energy
of the system is, to within an additive constant, $E[\phi ,R]=E_{dis}+E_{int}
$. We refer to this as the particle on a string model for the transport, as
the above expression is identical to the energy of a stretched string, which
is displaced from its equilibrium position by the particle, to form a
bubble. \ \ We assume overdamped, Langevin dynamics for the string and the
particle, given by $\left(
\begin{array}{cc}
\zeta  & 0 \\
0 & \zeta _{p}%
\end{array}%
\right) \frac{\partial }{\partial t}\Psi (x,R,t)=-\left(
\begin{array}{c}
\frac{\delta E[\phi ,R]}{\delta \phi (x)} \\
\frac{\partial E[\phi ,R]}{\partial R}%
\end{array}%
\right) +F(x,t)$

$\Psi (x,R,t)$ is $2\times 1$ column vector with elements $\phi (x,t)$ and $R
$. $\ \zeta $ is the friction coefficient for the string and $\zeta _{p}$
for the particle. We shall use\ $^{\dagger }$ to denote the transpose. $\
F(x,t)=\left( f(x,t),g(t)\right) ^{\dagger }$\ where $f(x,t)$ and $g(t)$ are
Gaussian random forces with mean zero and $\left\langle
f(x_{1},t_{1})f(x_{2},t_{2})\right\rangle =2\zeta k_{B}T\delta
(x_{1}-x_{2})\delta (t_{1}-t_{2})$ and $\left\langle
g(t_{1})g(t_{2})\right\rangle =2\zeta _{p}k_{B}T\delta (t_{1}-t_{2})$.\ \ We
define $\beta =1/k_{B}T$ and denote the time of relaxation of the long wave
length oscillations of the string by $\tau =\zeta /k$. \ \ A model with
similar structure has been studied in the context of DNA replication by
Bhattacharjee \cite{Bhattacharjee}. \

It is convenient to work with dimensionless quantities, and use the original
symbols themselves to denote them. \ Thus we change to $\beta E\rightarrow E,
$ $\beta E_{b}\rightarrow E_{b},$ $x/R_{0}\rightarrow x,$ $\phi
(x)/R_{0}\rightarrow \phi (x),$ $\alpha /R_{0}\rightarrow \alpha ,$ $%
R/R_{0}\rightarrow R,$ $\beta kR_{0}^{3}\rightarrow k,$ $\beta \Delta
kR_{0}^{3}\rightarrow \Delta k$, $\sigma \rightarrow \sigma \beta R_{0}$, $t/%
{\beta \zeta R_{0}^{3}}\rightarrow t$ and $\zeta _{p}/{\zeta R_{0}}%
\rightarrow \zeta _{p}$. \ The minimum energy configuration for the system
obeys the two equations $\frac{\delta E[\phi ,R]}{\delta \phi (x)}=0$ and $%
\frac{\partial E[\phi ,R]}{\partial R}=0$. \ These lead to $
\sigma \frac{\partial ^{2}\phi (x)}{\partial x^{2}}-k\phi (x)-\left( \Delta
k(\phi (y)-a\right) -k\alpha )Q(x-R)=0\notag$ 
and $\int_{-\infty }^{\infty }dxQ^{\prime }(x-R)({(k+\Delta k)\left\{ \phi
(x)-\alpha \right\} ^{2}-k\phi }^{2}{(x))}=0$. \ The prime, $^{\prime }$ is
used to denote differentiation with respect to $x$. \ As the shape of the
distortion would obviously depend on where the particle is located, it is
clear that the distortion would have the form $\phi _{c}(y)$ where $y=x-R$.
\ In terms of $y$, the first of the two equations become $\sigma \frac{%
\partial ^{2}\phi _{c}(y)}{\partial y^{2}}-k\phi _{c}(y)-\left( \Delta
k(\phi (y)-\alpha \right) -k\alpha )Q(y)=0$. The second equation is easily
satisfied by having $\phi _{c}(-y)=\phi _{c}(y)$. $\ $Once $\phi _{c}$
satisfying these conditions is found, it may be used to get the minimum
energy, $E_{c}=$ $E[\phi _{c}(x-R),R]$. \ The minimum energy configuration
with the particle stationary at $R_{b}$, and with the center of the
distortion coinciding with $R_{b}$ shall be referred to as having the bubble
at $R_{b}$ and this may be specified by the function  $\Psi
_{c}(x-R_{b},R_{b})=\left( \phi _{c}(x-R_{b}),R_{b}\right) ^{\dagger }$. \ \
Movement of the bubble as a whole is described as change of $\Psi
_{c}(x-R_{b},R_{b})$ by change of $R_{b}$. \ It is clear that this motion is
translational as the energy of the system is unchanged by this movement.
This causes the existence of a \textquotedblleft zero mode" \cite{rajaraman}%
.\ At a finite temperature, there would be fluctuations which will cause the
bubble to execute random motion. It is to be noted that the distortion and
the particle can fluctuate in opposite directions and therefore $R_{b}$ and $%
R$ follow different dynamics. Following the methods of instanton theory \cite%
{rajaraman}, that has been used in a similar context
\cite{kls,kls2}, we write the state of the system at any time as the
configuration with the bubble located at $R_{b}$ plus fluctuations
about this configuration, which we expand in terms of the normal
modes about the bubble at $R_{b}$. Thus, we write $\Psi (x,R,t)=\Psi
_{c}(x-R_{b}(t),R_{b}(t))+\sum\limits_{l=1}^{\infty }C_{l}(t)\Psi
_{l}(x-R_{b}(t))$, where $\Psi _{l}(x-R_{b})=(\phi
_{l}(x-R_{b}),J_{l})^{\dagger }$ are normal modes, the equation for
which will be defined later. We now substitute the above expansion
into the Langevin equation and expand the RHS up to first order in
$C_{l}(t)$. The
result is 
$\mathbf{\varsigma }\overset{.}{R}_{b}(-\phi _{c}^{\prime
}(x-R_{b}(t)),1)^{\dagger }+\sum\limits_{l=1}^{\infty }\overset{.}{C}%
_{l}(t)\Psi _{l}(x-R_{b}(t))\ =\sum\limits_{l=1}^{\infty }C_{l}(t)\left\{
\widehat{L}\Psi _{l}(x-R_{b}(t))-\overset{.}{R}_{b}\Psi _{l}^{\prime
}(x-R_{b}(t))\right\} +F(x,t)$
where $\mathbf{\varsigma }$ is a $2\times 2$ diagonal matrix with
diagonal elements $1$ and $\zeta _{p}$. The operator $\widehat{L}$
is defined by 
\begin{eqnarray*}
\widehat{L}\Psi _{l}(y)&=& \nonumber
\end{eqnarray*}
\begin{eqnarray*}
\left(
\begin{array}{c}
\left\{ -\sigma \partial ^{2}/\partial y^{2}+k+\Delta kQ(y)\right\}
\phi
_{l}(y)-Q^{\prime }(y)\left\{ \Delta {k(\phi _{c}(y)-\alpha )-\alpha k}%
\right\} J_{l} \\
-\int_{-\infty }^{\infty }Q^{\prime }(y^{\prime })\left\{ \Delta
{k(\phi _{c}(y}^{\prime }{)-\alpha )-\alpha k}\right\} \phi
_{l}(y^{\prime })+J_{l}\int_{-\infty }^{\infty }dyQ^{\prime \prime
}(y)\left\{ (k+\Delta
k)/2\left( \phi _{c}(y)-\alpha \right) ^{2}-(k/2)\phi _{c}(y)^{2}\right\}%
\end{array}
\right)
\end{eqnarray*}
We take $\Psi _{l}$ to obey the eigenvalue equation $\widehat{L}\Psi
_{l}=
\lambda _{l}\Psi _{l}$. It is convenient to introduce an inner-product $%
\left( \Psi _{l}|\Psi _{l_{1}}\right) =\left\langle \phi _{l}|\phi
_{l_{1}}\right\rangle +\zeta _{p}J_{l}^{\ast }J_{l_{1}}$, where $%
\left\langle \phi _{l}|\phi _{l_{1}}\right\rangle =$ $\int_{-\infty
}^{\infty }dx\phi _{l}^{\ast }(x)\phi _{l_{1}}(x)$. With this
definition of the inner product, $\widehat{L}$ is a hermitian
operator on the space spanned by $\Psi _{l}$. Further, $\partial
/\partial R_{b}\left( \phi _{c}(x-R_{b}),R_{b}\right) ^{\dagger
}=(-\phi _{c}^{\prime }(x-R_{b}),1)^{\dagger }$ is an eigen function
of $\widehat{L}$ with an
eigenvalue zero, as may be easily proved by differentiating $\left( \frac{%
\delta E[\phi ,R]}{\delta \phi (x)}\right) _{\phi _{c}(x-R_{b}),R_{b}}=0$
and $\left( \frac{\partial E[\phi ,R]}{\partial R}\right) _{\phi
_{c}(x-R_{b}),R_{b}}=0,$ \ with respect to $R_{b}$ and writing the results
in the matrix form. This is the zero mode of the system. After normalizing,
this mode may be written as $\Psi _{0}=(-\phi _{c}^{\prime
}(x-R_{b}),1)^{\dagger }/\sqrt{c}$, with $c=\zeta _{p}+\left\langle \phi
_{c}^{\prime }|\phi _{c}^{\prime }\right\rangle $. \ Note that in the spirit
of instanton approach the sum over $l$ in the expansion of $\Psi (x,R,t)$
does not include this mode. \ Further, the fact that $Q(y)$ is a symmetric
function, means that one can classify the eigenfunctions $\Psi _{l}$ based
upon the symmetry or antisymmetry of $\phi _{l}(y)$. \ If $\phi _{l}(y)$ is
symmetric, then $\Psi _{l}$ has the simple form $\Psi _{l}=(\chi
_{l},0)^{\dagger }$, where $\chi _{l}$ is a symmetric eigenfunction of the
operator $\left( -\sigma \frac{\partial ^{2}}{\partial y^{2}}+k+\Delta
kQ(y)\right) $. \ Further, for this case, $\lambda _{l}=\varepsilon _{l}$,
where $\varepsilon _{l}$ is the eigenvalue associated with $\chi _{l}$, a
fact we will use later. \

On taking inner product of the equation involving $C_{l}(t)$ with $\Psi
_{0}(x-R_{b}(t))$ we get $\overset{.}{R}_{b}(t)$ $=h(t)+(\dot{R}%
_{b}(t)/c)\sum\limits_{l=1}^{\infty }C_{l}(t)\left\langle \phi
_{c}^{^{\prime \prime }}|\phi _{l}\right\rangle $, where $h(t)=$ $%
(g(t)-\left\langle \phi _{c}^{^{\prime }}\mid f(t)\right\rangle )/c$. \ To
get the simplest approximation for the diffusion coefficient for the bubble,
we neglect the \textquotedblleft string-phonon-particle scattering" (SPP)\
term (the last term) of the above equation involving $\dot{R}_{b}(t)$ and
get $\dot{R_{b}}(t)=h(t)$. \ The diffusion coefficient is then found to be $%
D_{0}=\frac{1}{2{t}}\int_{0}^{t}dt_{1}\int_{0}^{t}dt_{2}\left\langle
h(t_{1})|h(t_{2})\right\rangle =c^{-1}$. One can write $D_{0}$ as a harmonic
mean $D_{0}=D_{p}D_{s}\left( D_{p}+D_{s}\right) ^{-1},$with $D_{p}=\zeta
_{p}^{-1}$, $D_{s}=\left\langle \phi _{c}^{^{\prime }}\mid \phi
_{c}^{^{\prime }}\right\rangle ^{-1}$. It is interesting that within this
approximation, $D_{0}$ is always less than $D_{p}$ implying that the
diffusion coefficient within the plug is less than outside, as expected \cite%
{ribbeck}. To get a better approximation, one has to include the last term
of the equation involving $\dot{R}_{b}(t)$. \ For this, we have to find $%
C_{l}(t)$. \ On taking inner-product of the equation involving $C_{l}(t)$
with $\Psi _{l}(x-R_{b}(t))$, and neglecting SPP scattering, we get $\overset%
{.}{C_{l}}(t)=-\lambda _{l}C_{l}(t)+\left( \Psi _{l}\mid F(t)\right) $,
which can be solved with the condition $C_{l}(-\infty )=0$ to get $%
C_{l}(t)=\int_{-\infty }^{t}dt_{1}\left( \Psi _{l}\mid F(t_{1})\right) \exp
(-\lambda _{l}(t-t_{1}))$. One should now solve for $\dot{R_{b}}(t)$ and
then calculate the diffusion coefficient. However, this is difficult, as the
noise is multiplicative. We therefore adopt a simple minded approach in
which $\dot{R_{b}}(t)$ on rhs is replaced by $h(t)$ to get the following
equation $\dot{R_{b}}(t)=h(t)\left( 1+B(t)/c\right) $, where $%
B(t)=\sum_{l=1}^{\infty }\int_{-\infty }^{t}dt_{1}\left( \Psi _{l}\mid
F(t_{1})\right) \exp (-\lambda _{l}(t-t_{1}))$ $\left\langle \phi
_{c}^{^{\prime \prime }}\mid \phi _{l}\right\rangle $. If one further
assumes that $h(t)$ and $B(t)$ are uncorrelated then the diffusion
coefficient becomes $D=D_{0}+\frac{1}{2tc^{2}}\int_{0}^{t}dt_{1}%
\int_{0}^{t}dt_{2}\left\langle h(t_{1})h(t_{2})\right\rangle \left\langle
B(t_{1})B(t_{2})\right\rangle $. This is easy to calculate since the noises
are delta function correlated and it becomes, $D=D_{0}\left(
1+B/c^{2}\right) $, where $B=\left\langle B(t)B(t)\right\rangle =\sum_{l\neq
0}\left\vert \left\langle \phi _{c}^{^{\prime \prime }}\mid \phi
_{l}\right\rangle \right\vert ^{2}/\lambda _{l}$. As $\phi _{c}^{^{\prime
\prime }}(x)$ is an even function of $x$, only $\phi _{l}$ that are
symmetric will contribute in this sum. \ But as we saw above, they are
identical with $\chi _{l}$, and for them $\lambda _{l}=\varepsilon _{l}$. \
We can therefore write $B=\sum_{l}\left\vert \left\langle \phi
_{c}^{^{\prime \prime }}\mid \chi _{l}\right\rangle \right\vert
^{2}/\varepsilon _{l}$. \ It is clear that one can expand the sum over $l$
to include all $l$. \ Thus we get $D=D_{0}\left( 1+\left\langle \phi
_{c}^{^{\prime \prime }}\left\vert \widehat{G}\right\vert \phi
_{c}^{^{\prime \prime }}\right\rangle /c^{2}\right) .$ $\widehat{G}=\left(
-\sigma \frac{\partial ^{2}}{\partial y^{2}}+k+\Delta kQ(y)\right) ^{-1}$ is
the Green's operator. Note that in the Schr\"{o}dinger operator $-\sigma
\frac{\partial ^{2}}{\partial y^{2}}+k+\Delta kQ(y)$, the term $\Delta kQ(y)$
vanishes at infinity. \ It is straightforward to calculate the diffusion
coefficient for simple models of $Q(y)$. \ We define an enhancement factor  $%
r=D/D_{0}=\left( 1+\left\langle \phi _{c}^{^{\prime \prime }}\left\vert
\widehat{G}\right\vert \phi _{c}^{^{\prime \prime }}\right\rangle
/c^{2}\right) $. 
\begin{figure}[tbp]
\centering \epsfig{file=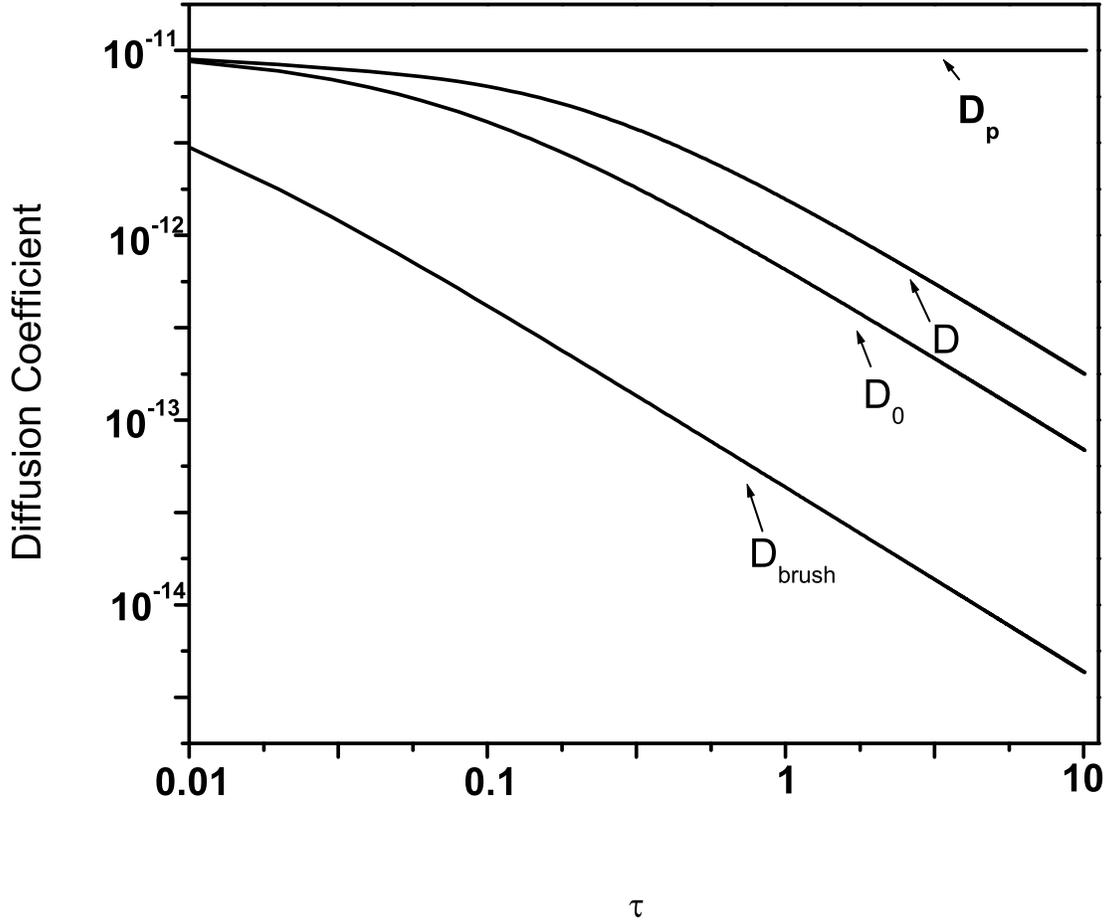,width=\linewidth}
\caption{Plot of the diffusion coefficient of the macromolecule against the
longest relaxation time ($\protect\tau $). $D_{p}$ is the diffusion
coefficient outside the gel. $D_{0}$ and $D$ with $\protect\sigma =2\times
10^{-13}J/m$ (for the gel) and $D_{brush}$ with $\protect\sigma =2\times
10^{-15}J/m$. }
\label{diffusion2}
\end{figure}
We now perform calculations for a simple model with $Q(y)=\theta
(y-1)-\theta (y+1)$. Further, we put $\Delta k=0$ so that the calculation
becomes easy. Then one can find $\phi _{c}$ easily and obtain $E_{c}=$ $%
-E_{b}+{\alpha }^{2}\sqrt{k\sigma }\sinh (\sqrt{k/\sigma })\exp (-\sqrt{%
k/\sigma })$, where $E_{b}$ is the dimensionless value of the binding energy
($=E_{b}\beta $). If $E_{c}\leq 0$, then the entry into the NPC has no
activation energy (however, exit would be activated). The Green's function
is $G(y,y_{1})=\left( \sqrt{k/4\sigma }\right) \exp (-\sqrt{\frac{k}{\sigma }%
}\left\vert y-y_{1}\right\vert )$ so that $\left\langle \phi _{c}^{^{\prime
\prime }}\left\vert \widehat{G}\right\vert \phi _{c}^{^{\prime \prime
}}\right\rangle =\left( \sqrt{k/\sigma }\right) $ $\left( 1-\exp (-2\sqrt{%
k/\sigma })-2\sqrt{k/\sigma }+4k/\sigma \right) /8$. \ With these, we get $%
D_{0}=2\sigma /\delta $ with $\delta =$ $(2\sigma \zeta _{p}+\alpha ^{2}(%
\sqrt{k\sigma }-\exp (-2\sqrt{\frac{k}{\sigma }})(2k+\sqrt{k\sigma })))$ and
$r=1+$ $\frac{\alpha ^{2}\sqrt{\frac{k}{\sigma }}\exp (2\sqrt{\frac{k}{%
\sigma }})\left( -1+\exp (2\sqrt{\frac{k}{\sigma }})-2\sqrt{\frac{k}{\sigma }%
}+4\frac{k}{\sigma }\right) }{2\left( (-1+\exp (2\sqrt{\frac{k}{\sigma }}%
))\alpha ^{2}\sqrt{\frac{k}{\sigma }}-\frac{2k\alpha ^{2}}{\sigma }+2\zeta
_{p}\exp (2\sqrt{\frac{k}{\sigma }})\right) ^{2}}$.

We now estimate the diffusion coefficient of the protein, assuming that the
plug is a gel. \ As experimental information on the properties of the gel is
scarce, the numbers used are only rough estimates. \ As the protein has size
of a few nanometers, we take $R_{0}=10$ $nm$, $\alpha =10$ $nm$ and $%
L_{Y}=L_{Z}=30$ $nm$. Temperature is taken to be $300$ K . \ We take the
diffusion coefficient of the protein outside the network to be $%
D_{p}=k_{B}T/\zeta _{p}=10^{-11}m^{2}s^{-1}$, which gives the value of $%
\zeta _{p}=4.14\times 10^{-10}Js/m^{2}$. \ Considering a slab of dimensions $%
L_{X},L_{Y}$ and $L_{Z}(=L_{Y})$, the energy required to deform it by a
constant amount $a$ in the Z-direction is $L_{X}a^{2}Y/2,$where $Y$ is the
Young's modulus. \ Using our equation for $E_{dis}$ for the same situation
gives $kL_{X}a^{2}/2$. \ Equating the two, we get $k=Y$. \ \ $Y$ for the
hydrogel formed by nucleoporins in vitro has been measured to be $2000$ $Pa$
\cite{frey}. \ \ With this value for $k$, to produce a deformation with $%
\phi =10$ $nm$ over a length of $20$ $nm$ requires only an energy of $%
1.2kJ/mol$, which is $\sim k_{B}T/2$. \ So the deformation of the gel to
produce a hole of the required size requires only thermal energy. \ So then,
why does not the particle just deform the gel and get inside? \ The answer
must be that to get inside, it will have to break the network. \ It is known
that a colloidal particle of size $5$ $nm$ $\ $is not able to get into the
pore \cite{paine, bickel}. \ This means that there must be at least one
hydrophobic contact to be broken to create a hole of volume $(5nm)^{3}$. We
will take the energy of a hydrophobic contact as $5k_{B}T$. Therefore, to
create a hole of the size of a protein, ($20nm\times 10nm\times 10nm$), one
will have to break $16$ hydrophobic contacts which will require an energy of
$80k_{B}T$. This rather large energy requirement can be compensated by the
formation of roughly the same number of hydrophobic contacts between the
protein and the network. \ \ We estimated $\sigma $ by putting the condition
that the deformation energy due to the two quadratic terms of $E_{dis}$ have
the same value for creation of a deformation of size $10$ $nm$ over a length
of $20$ $nm$. \ This gave $\sigma $ for the gel $\sigma _{gel}=2\times
10^{-13}J/m$. As no experimental data on the relaxation time of the network
is available, we took $\zeta =2$ $Js/m^{3},$ so that the long wave length
relaxation time $\tau $ of the gel would be $1$ $ms$. This seems reasonable
as the gel is formed due to rather weak hydrophobic interactions among FG
units. \ These data correspond to the following values for the dimensionless
parameters: $\alpha =1,k=1,\zeta _{p}=0.0207,D_{p}=1/\zeta _{p}=48.3092$.
With these values, we find: \ The value of $D_{0}$, diffusion coefficient of
the protein inside the network to be $6.51\times 10^{-13}m^{2}/s$. \ This
means that putting into the network has reduced the diffusion coefficient by
roughly a factor of $\sim 15$, as expected\cite{bickel}. \ The value of $%
r=2.4$ and hence $D=1.56\times 10^{-12}m^{2}/s$ - noise in the gel enhances
the diffusion considerably. Putting the length of the NPC to be $50nm$, this
would give a residence time of $1.6$ $ms$ for the particle. This estimates a
transport rate of $\sim 600$ proteins/second which is roughly in agreement
with the experiments \cite{ribbeck}. In Fig. \ref{diffusion2} we have made a
plot of diffusion coefficient against the relaxation time ($\tau $) of the
gel, keeping all other parameters fixed which shows that the faster is the
relaxation of the gel, faster is the diffusion of the macromolecule
(protein) inside it. \ Further, the enhancement factor, $r$ is $\sim 2$ for $%
\tau >1$ $ms$.

What would happen if the plug were a brush? \ In a brush, there are no
hydrophobic contacts and hence there is no network. \ \ Any contact would
contribute to the elasticity of the plug and therefore, if the plug were a
brush, the value of $\sigma $ would be much lower. \ \ The model is easily
analyzed in the limit $\sigma \rightarrow 0$, to find that $D_{0}\rightarrow
0$, $r\rightarrow 1$ so that $D\rightarrow 0$. \ Thus the diffusion
coefficient of the particle within the brush would be much lower than in the
gel. \ To demonstrate this, we have calculated the diffusion coefficient for
the case with $\sigma =\sigma _{gel}/100$ and the results are given in the
Fig. \ref{diffusion2}, and it is seen that the diffusion coefficient is
lowered by a factor of $1/35$. \

It is also interesting to compare our results with those of Zilman \emph{et
al} \cite{Zilman2007}. \ In their model, the greater binding energy of the
particle to the FG units would enhance the transport, by retaining the
particles longer in the pore. \ The same thing would happen within our model
too. \ The two models are similar in that the movement of the particle is
co-operative, with the contacts at the front being formed at the same time
as the contacts at the back are being broken, leading to easy motion of the
particle within the NPC. \ However, over approach is more detailed, and is
able to take care of the limit where $\sigma $ is small, which as we have
argued, is very important.

In summary, we have proposed a model for the transport of a protein
through the NPC. We find that a reversible gel would lead to larger
diffusion coefficients than the polymer brush and hence we suggest
that the plug of the NPC is a gel. However,  final confirmation of
this needs more experimental work and simulations.

 R. Chakrabarti's work was supported by CSIR (India).

\end{document}